# Electromagnetically induced gratings created by extremely short non-overlapping pulses of light in a three-level resonant medium

Rostislav Arkhipov [1,2]

[1]St. Petersburg State University, Saint Petersburg, Russian Federation,

[2]Ioffe Institute, Saint Petersburg, Russian Federation.

e-mail: arkhipovrostislav@gmail.com

**Keywords**

Half-cycle pulses, attosecond pulses, unipolar pulses, electromagnetically induced gratings, polarization waves, spatiotemporal photonic crystals

**Abstract**

In a fixed spectral range, single- and half-cycle electromagnetic pulses have the shortest duration. Half-cycle pulses are promising tools for ultrafast control of quantum systems. Previously, the possibility of using a sequence of single- and half-cycle attosecond pulses to generate and ultrafast control light-induced population difference gratings has been demonstrated. However, such studies have been carried out using different approximations. For example, when the medium is modelled in the two-level approximation. In this paper, based on the numerical solution of the system of Maxwell-Bloch equations, it is shown that it is possible to generate and control population gratings in a three-level medium without using the approximations used in previous studies. It is shown that taking into account the additional level of the medium does not lead to a violation of the effect of generating such gratings. This extends the applicability of previous results.

1. Introduction

The generation of ultrashort electromagnetic pulses lasting several field periods is one of the main topics of modern optics in recent decades [1-4]. These pulses are actively used to study electron dynamics in media at femtosecond and attosecond timescales [5-10]. The relevance of this topic is underlined by the recent award of the Nobel Prize in Physics in the field of the generation of attosecond pulses [11].

Conventional ultra-short pulses contain a number of half-waves of the electric field strength [1-10]. And to shorten the duration of the pulses, one has to go from the visible to the ultraviolet and even the X-ray part of the spectrum, i.e. to the higher frequency range. This is used to obtain attosecond pulses in the high-order harmonic generation method [1-10]. However, it is also possible to use another method.

The shortest duration in a given spectral range is obtained by removing all the half-waves from the original multi-cycle pulse and leaving only one. In this case, a unipolar half-cycle pulse is obtained [12-18]. For such pulses, the electrical area plays an important role. It is defined as the integral of the field strength $E(t)$ over time $t$ at a given point in space, $S_E = \int E(t)dt$ [13-16]. This value is always close to zero for ordinary bipolar pulses. It is therefore meaningless and has long been ignored.

There is an interest in obtaining such pulses and this has been actively developed in recent times [12-18]. On the one hand, such pulses have a number of unusual properties. On the other hand, they have a very fast and unidirectional effect on charges, which opens up a wide range of possibilities for their application, from the effective excitation of atoms, molecules and nanostructures [19-22] to the acceleration of charges [23] and even holography with ultra-high temporal resolution of fast-moving objects [24].

Standard theoretical approaches (Keldysh theory and related formalisms) become invalid when the duration of such pulses is shorter than the orbital period of the electron in the atom [25]. And new ones should apply. The result of this study showed that in this case the excitation of the bound state probability is determined by the electrical pulse area rather than its energy [19-22].

We note that, in spite of the complexity of obtaining strictly unipolar pulses, there is a large number of works that consider the possibility of obtaining pulses that contain a characteristic burst (half-wave) of a field of the same polarity [6,26-35] and weak leading edge of the opposite polarity. Reviews [11-17] and cited literature summarize the results of these studies.

Another possible application of ultra-short [36-39] and extremely short pulses [40-51] is the creation and control of electromagnetically induced atomic population gratings (EMIGs). In conventional approaches, such EMIGs are created by the superposition of two coherent monochromatic laser beams in the medium by the formation of the interference pattern [52]. Such

EMIGs, created in this way, have a wide range of applications in the various fields of optics [53-57]. In particular, light diffraction from such EMIGs is used to determine medium susceptibilities or relaxation times [52-57]. In addition, such gratings created in optical fibers are used as high-efficiency reflectors (fiber Bragg gratings) [58-61].

However, the conventional approach does not allow for the rapid control of such gratings [49]. In order to do this, extremely short pulses can be used. The shorter the pulses, the faster the grating can be driven. Therefore, for the creation and ultrafast control of such EMIGs, single-cycle and half-cycle pulses can be an attractive tool [39-48]. Direct interference is impossible with single- and half-cycle pulses. However, gratings can be created if the pulses interact coherently with the medium, i.e. the pulse duration and the delay between pulses are shorter than the medium polarization relaxation time $T_2$. This can be achieved if the pulses do not overlap or collide with each other in the medium.

The interference of the polarization waves of the medium that generated each previous pulse with the subsequent pulses is the physical mechanism for the generation of such gratings [40-51]. In the case of half-cycle pulses, when their direct interference is impossible, when the amplitude of the pulses is small, their effect on quantum systems can be described in terms of interference of pulse areas [17,47,49]. In this case, it can be said that the creation of gratings by half-cycle pulses is due to the interference of the electrical areas of the pulses. Early research in this direction is reviewed in Ref. [17,46] and cited literature.

The possibility of guidance and ultra-fast control of such EMIGs by means of single-cycle and half-cycle pulses has been theoretically studied in the works [17,40-51]. However, these studies have been carried out in a variety of approximations. For example, the resonant medium was modelled using the simplest two-level approximation [40-43]. In the case of half-cycle pulses, however, they interact non-resonantly with the medium. The two-level approximation remains valid for not very large values of their electrical area according to results of Ref. [62].

For the consideration of additional levels of the medium, an approximate solution of the time-dependent Schrödinger equation (TDSE) was used in the approximation of a small amplitude of the field excitation of the medium [21,43-44,49]. In Ref. [48] numerical solution of TDSE was used to study gratings. In these studies, the medium was considered to be rarefied. And the problem of pulse propagation in an extended medium was reduced to the effect of a sequence of pulses with variable delay on a single quantum system. The effects of coherent propagation of pulses in an extended medium, the dynamics of polarization waves and their influence on the formation of gratings are obviously not considered in this approach.

Taking into account the shortcomings of previous studies, this work theoretically investigates the possibility of ultrafast guidance and control of population gratings by means of a sequence of extremely short pulses (single- and half-cycle ones) in a three-level resonant medium. The analysis is based on a numerical solution of the system of Maxwell-Bloch equations, taking into account the effects of pulse propagation in the medium. The possibility of creating gratings and controlling them at ultrafast speeds is demonstrated. This justifies the approximations used previously and extends the scope of the results obtained.

## 2. Methods

### 2.1. The theoretical model

The interaction of a three-level medium with the driving field of the pulses can be described by the following system of Maxwell-Bloch equations. It includes the system of equations for the elements of the density matrix of the three-level system coupled to the 1D wave equation for the electric field strength. This system is written without both the slowly varying envelope and the rotating wave approximations [49,63]:

$$\frac{\partial}{\partial t}\rho_{21} = -\rho_{21}/T_{21} - -i\omega_{12}\rho_{21} - i\frac{d_{12}}{\hbar}E(\rho_{22} - \rho_{11}) - i\frac{d_{13}}{\hbar}E\rho_{23} + i\frac{d_{23}}{\hbar}E\rho_{31}, \quad (1)$$

$$\frac{\partial}{\partial t}\rho_{32} = -\rho_{32}/T_{32} - i\omega_{32}\rho_{32} - i\frac{d_{23}}{\hbar}E(\rho_{33} - \rho_{22}) - i\frac{d_{12}}{\hbar}E\rho_{31} + i\frac{d_{13}}{\hbar}E\rho_{21}, \quad (2)$$

$$\frac{\partial}{\partial t}\rho_{31} = -\rho_{31}/T_{31} - i\omega_{31}\rho_{31} - i\frac{d_{13}}{\hbar}E(\rho_{33} - \rho_{11}) - i\frac{d_{12}}{\hbar}E\rho_{32} + i\frac{d_{23}}{\hbar}E\rho_{21}, \quad (3)$$

$$\frac{\partial}{\partial t}\rho_{11} = -\frac{\rho_{22}}{T_{22}} + \frac{\rho_{33}}{T_{33}} + i\frac{d_{12}}{\hbar}E(\rho_{21} - \rho_{21}^*) - i\frac{d_{13}}{\hbar}E(\rho_{13} - \rho_{13}^*), \quad (4)$$

$$\frac{\partial}{\partial t}\rho_{22} = -\rho_{22}/T_{22} - i\frac{d_{12}}{\hbar}E(\rho_{21} - \rho_{21}^*) - i\frac{d_{23}}{\hbar}E(\rho_{23} - \rho_{23}^*), \quad (5)$$

$$\frac{\partial}{\partial t}\rho_{33} = -\frac{\rho_{33}}{T_{33}} + i\frac{d_{13}}{\hbar}E(\rho_{13} - \rho_{13}^*) + i\frac{d_{23}}{\hbar}E(\rho_{23} - \rho_{23}^*). \quad (6)$$

$$P(z,t) = 2N_0 d_{12} Re\rho_{12}(z,t) + 2N_0 d_{13} Re\rho_{13}(z,t) + 2N_0 d_{12} Re\rho_{12}(z,t) + 2N_0 d_{23} Re\rho_{32}(z,t).$$
(7)

$$\frac{\partial^2 E(z,t)}{\partial z^2} - \frac{1}{c^2}\frac{\partial^2 E(z,t)}{\partial t^2} = \frac{4\pi}{c^2}\frac{\partial^2 P(z,t)}{\partial t^2}. \qquad (8)$$

Equations (1)-(3) describe the evolution of off-diagonal elements of the density matrix $\rho_{21}$, $\rho_{32}$, $\rho_{31}$, which determine the dynamics of the medium polarization $P(z,t)$ governed by Eq. (7). Equations (4) to (6) describe the evolution of the diagonal elements of the density matrix $\rho_{11}$, $\rho_{22}$, $\rho_{33}$. These elements represent the populations of the first, second and third states of the medium, respectively. Other parameters are: $\omega_{21}$, $\omega_{32}$, $\omega_{31}$ – transition frequencies of the medium, $d_{12}$, $d_{13}$, $d_{23}$ – transition matrix dipole moments, $T_{ik}$ are the relaxation times.

The total length of the computational domain is $L = 12\lambda_0$. The medium was located between the points $z_1 = 4.2\lambda_0$ and $z_2 = 7.8\lambda_0$. In order to generate the pulse train, we used the zero boundary conditions: by reaching the left and right boundaries of the computational domain, the pulses reflected from it and moved back into the medium along the z-axis. In the following subsections the results of the numerical simulations are presented

### 3. Results

#### 3.1. Grating formation by single-cycle non overlapping pulses

Pairs of one-cycle pulses 1 and 2 were injected into the medium from left/right and right/left, respectively, see Fig.1

$$E(z = 0, t) = E_{01} e^{-\frac{(t-\Delta_1)^2}{\tau^2}} \sin[\omega_0(t - \Delta_1)], \quad (9)$$

$$E(z = L, t) = E_{02} e^{-\frac{(t-\Delta_2)^2}{\tau^2}} \sin[\omega_0(t - \Delta_2)], \quad (10)$$

Delays $\Delta_{1,2}$ were chosen in a such away to avoid simulataneously pulse overlap in the medium.

The parameters given in Table were used to model the medium in the 3-level approximation. The parameters (transition frequencies and dipole moments) close to Rb[87] [64] were used for simplicity. Systems of density matrix equations was solved using the 4th order Runge-Kutta method. Wave equation was solved using standard finite-difference method described in [65]. The parameters of calculations are given in the Table.

**Table. Parameters used in our simulations**

| Parameter | Its value |
|---|---|
| amplitude of the pulse | $E_0 = 200\,000$ ESU |
| pulse duration | $\tau = 440$ as |
| Delay $\Delta_1$ | $\Delta_1 = 2.5\tau$ |
| Delay $\Delta_2$ | $\Delta_2 = 33.5\tau$ |
| 1→2 transition wavelength frequency $\omega_{12}$ | $\lambda_{12} = \lambda_0 = 794.7$ nm $\omega_{12} = 2.37 \cdot 10^{15}$ rad/s |
| dipole moment, transition 1→2 | $d_{12} = 7.6$ D |
| 1→3 transition wavelength frequency $\omega_{13}$ | $\lambda_{13} = 780$ nm $\omega_{13} = 2.41 \cdot 10^{15}$ rad/s |
| dipole moment, transition 1→3 | $d_{13} = 10.7$ D |
| 2→3 transition wavelength frequency $\omega_{23}$ | $\omega_{23} = \omega_{13} - \omega_{12}$ |
| dipole moment, transition 2→3 | $d_{23} = 0$ D |
| Medium density | $N_0 = 10^{14}$ cm$^{-3}$ |

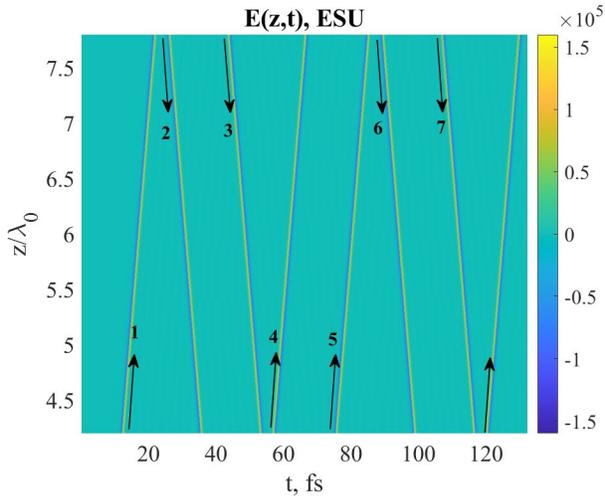

**Fig. 1**. The diagram shows the direction of propagation of single cycle pulses in the medium. The arrows indicate the direction of propagation and the numbers indicate the number of pulses.

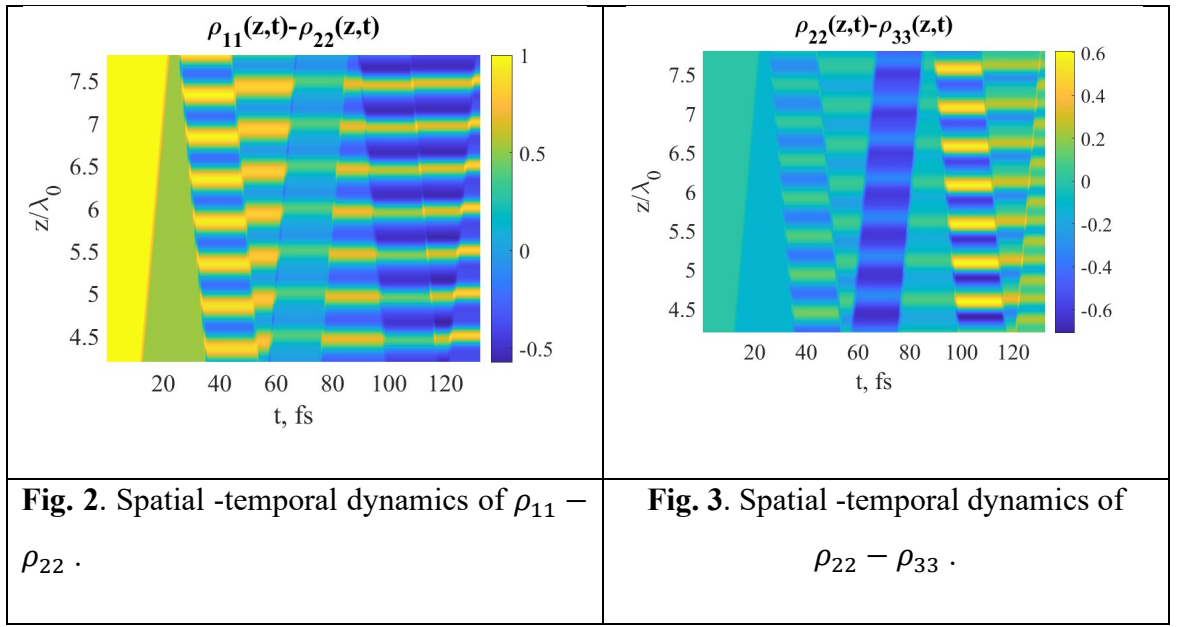

| | |
|---|---|
| **Fig. 2**. Spatial -temporal dynamics of $\rho_{11} - \rho_{22}$. | **Fig. 3**. Spatial -temporal dynamics of $\rho_{22} - \rho_{33}$. |

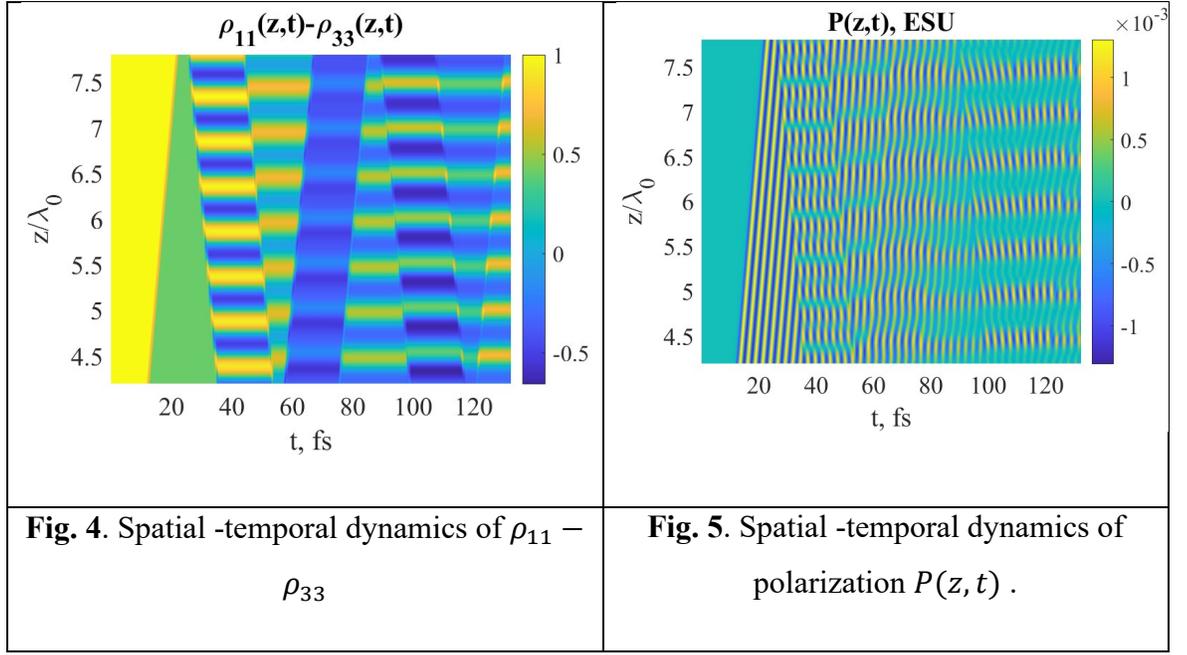

**Fig. 4**. Spatial-temporal dynamics of $\rho_{11} - \rho_{33}$

**Fig. 5**. Spatial-temporal dynamics of polarization $P(z,t)$.

Figures 2-4 illustrate the spatial and temporal dynamics of the population difference at different resonance transitions of the three-level medium. Fig.5 shows the dynamics of the polarization of the medium. These figures show the possibility of guiding the population grating at each transition of the medium after the second pulse. After the 3rd and 4th pulses the gratings are controlled, i.e. the gratings modulation depth is changed. Both running and standing polarization waves can be seen in Figure 5. In calculations for a two-level medium, similar results were observed [40-42]. Demonstrating the possibility of grating formation in a multilevel medium is the main purpose of these calculations. To study the dynamics of the grating depending on the problem parameters is beyond the scope of this paper.

### 3.2. Grating formation by half-cycle non overlapping pulses

In this subsection, in the numerical calculations, pairs of half-cycle unipolar pulses 1 and 2 were injected into the medium from left/right and right/left, respectively, similar to that in Fig.1

$$E(z=0,t) = E_{01} e^{-\frac{(t-\Delta_1)^2}{\tau^2}}, \quad (11)$$

$$E(z=L,t) = E_{02} e^{-\frac{(t-\Delta_2)^2}{\tau^2}}. \quad (12)$$

Delays also $\Delta_{1,2}$ were chosen in a such a way to avoid simultaneously pulse overlap in the medium. In this example $\Delta_1 = 2.5\tau, \Delta_2 = 30.5\tau$. Other parameters are the same as in Table. The scheme of the pulse action is also the same as in Fig.1.

We consider a completely unipolar pulse, neglecting the presence of a trailing edge of opposite polarity. Such an approximation is justified when this front is long and has a small amplitude. In this case it has no influence on the system. The significant influence is exerted by the field half-wave [21].

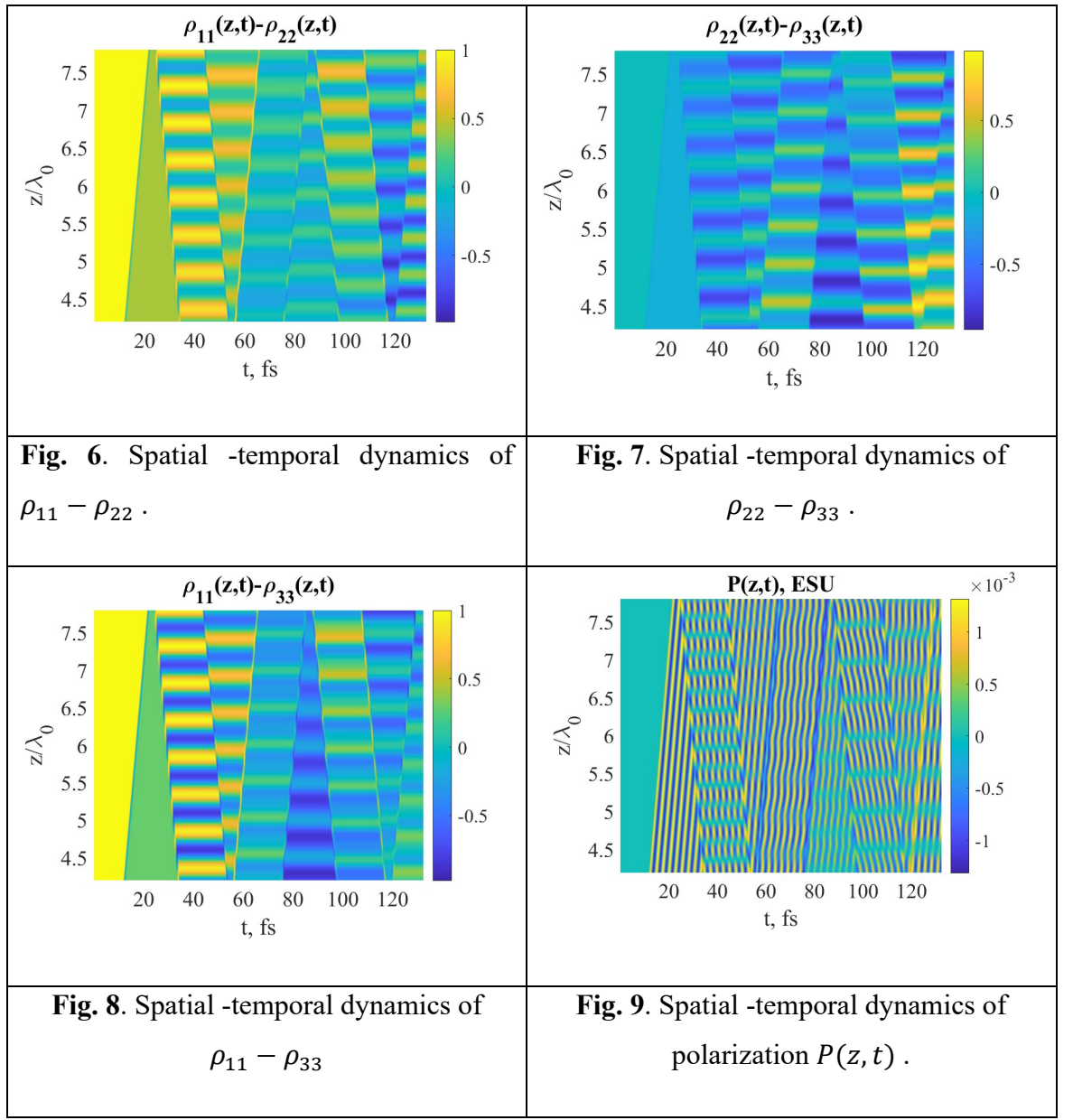

**Fig. 6**. Spatial-temporal dynamics of $\rho_{11} - \rho_{22}$.

**Fig. 7**. Spatial-temporal dynamics of $\rho_{22} - \rho_{33}$.

**Fig. 8**. Spatial-temporal dynamics of $\rho_{11} - \rho_{33}$

**Fig. 9**. Spatial-temporal dynamics of polarization $P(z,t)$.

Figures 6-8 illustrate the spatial and temporal dynamics of the population difference at different resonance transitions of the medium. Fig.9 shows the dynamics of the polarization of the medium. It can be seen that half-cycle pulses are also able to generate and control the gratings and that the third level does not suppress this phenomenon. The formation of the gratings is similar to that which has been observed previously in the two-level medium.

The physical scenario of grating formation is similar to that occurring in two-level medium [40-42]. The 1st pulse generates a polarization wave of the medium with different transition frequencies. The second pulse controls this wave coherently. A grating is formed at each resonant transition. The following pulses control the polarization oscillations that were created by the previous pulses. The result is the formation of different complex gratings, see Figs. 2-4 and 6-8. From these results, when more transitions are involved, the formation of gratings in a multilevel medium can be expected.

4. **Discussion**

The effect of EMIGs formation and ultrafast control by a sequence of extremely short attosecond pulses was clearly demonstrated by the results of the numerical simulations presented above. These results are in good agreement with the previous studies. These studies used different approximations. The three-level model therefore qualitatively reproduces the previous results obtained in the two-level model. The applicability of the two-level model for extremely short pulses is broadened by this fact.

This result does not surprise. Grating formation is based on the theoretically predicted carrier-wave Rabi-flopping effect in a two-level medium [66]. Subsequently, there has been experimental confirmation of this effect in a complex semiconductor system [67]. The results obtained using the two-level model are very good compared to experimental data. Thus, the results obtained using the two-level approximation are qualitatively reproduced in the multi-level system, which favors it for studying coherent light-matter interactions with extremely short pulses.

An interesting polarization wave behavior can be seen in Figures 5 and 9 at time intervals between approximately 60 and 80 fs. The polarization behavior shows a zigzagging structure. This behavior can be interpreted as the formation of narrow regions in the medium. In these regions polarization waves propagating in opposite directions are formed. Previously, in numerical calculations in a two-level model, when the pulses overlapped in the medium, similar structures were observed [68].

The three-level medium considered in this paper can be called a spatio-temporal photonic crystal according to modern terminology [69]. In fact, periodic gratings of populations are induced and modified by fast pulses. The result is a change in the refractive index of the medium, both in space and in time. Recently, the interest in studying such media has become more intense. In this sense, the studies above are a contribution to this new development in physics [70].

## 5. Conclusion

In conclusion, we have theoretically demonstrated the creation and control of population density gratings in a three-level medium by a train of non-overlapping in the medium of extremely short pulses using the numerical solution of the Maxwell-Bloch equations. This result is of fundamental importance. It confirms the existence of gratings, which were previously predicted using different simplifications, such as the two-level model.

In this way, the scope of application of the previously obtained results is extended. In addition, it opens up new possibilities for using single and half-cycle pulses to control the properties of the medium at ultrafast speeds. For example, such a grating could be used as a Bragg mirrors. Besides their application in holography with ultrahigh resolution of the fast-moving objects was shown in [24]. The advantage of the approach described above lies precisely in its ability to control ultrafast gratings using extremely short pulses which do not overlap.

**Acknowledgements**

This work was supported by the Russian Science Foundation, project no. 23-12-00012.

The author would like to thank his colleagues N.N. Rosanov, M.V. Arkhipov and A.V. Pakhomov for the helpful discussions.